\documentclass[12pt]{iopart}

\pdfoutput=1
 \expandafter\let\csname equation*\endcsname\relax
  \expandafter\let\csname endequation*\endcsname\relax
\usepackage{graphicx,bbm,color,amsmath}
\usepackage{dsfont, eufrak}

\let\mathfrak\undefined
\usepackage{amsfonts}
\usepackage{dsfont}
\usepackage{calligra}
\usepackage{calrsfs}
\usepackage[mathscr]{euscript}
\usepackage{hyperref}
\usepackage{cleveref}
\crefname{equation}{equation}{equations}
\Crefname{equation}{Equation}{Equations}
\crefrangelabelformat{equation}{(#3#1#4--#5#2#6)}

\crefmultiformat{equation}{equations (#2#1#3}{, #2#1#3)}{#2#1#3}{#2#1#3}
\Crefmultiformat{equation}{Equations (#2#1#3}{, #2#1#3)}{#2#1#3}{#2#1#3}

\newcommand{\be}{\begin{equation}}
\newcommand{\ee}{\end{equation}}

\newcommand{\bra}[1]{{\langle #1 \vert}}

\newcommand{\ket}[1]{{\vert #1 \rangle}}

\newcommand{\ave}[1]{{\langle #1\rangle}}
\newcommand{\ii}{ {\rm i} }
\newcommand{\dd}{ {\rm d} }

\newcommand{\bb}{ {\hat b} }

\newcommand{\y}{{\rm y}}
\newcommand{\x}{{\rm x}}
\newcommand{\z}{{\rm z}}

\newcommand{\half}{\frac{1}{2}}

\def\tr{{{\rm tr}}}
\def\ad{{\,{\rm ad}\,}}
\def\one{\mathbbm{1}}

\def\DD{\hat{\cal D}}
\newcommand{\La}{{\mathtt L}}
\newcommand{\Ra}{{\mathtt R}}
\newcommand{\LL}{{\hat{\cal L}}}
\newcommand{\lld}{{\hat{\ell}}}

\def\bb#1{\mathbf{#1}}

\def\End{{\,{\rm End}\,}}

\begin{document}

\title[Connected correlations, fluctuations and current of magnetization in the steady state...]{Connected correlations, fluctuations and current of magnetization in the steady state of boundary driven XXZ spin chains} 

\author{B. Bu\v ca$^1$ and T. Prosen$^2$}

\address{$^1$Department of Medical Physics and Biophysics, University of Split School of Medicine, 21000 Split, Croatia}
\address{$^2$Department of Physics, Faculty of Mathematics and Physics, University of Ljubljana, 1000 Ljubljana, Slovenia}

\begin{abstract}
We show how to exploit algebraic relations among the operators (or matrices) which constitute the non-equilibrium matrix product steady state of a boundary driven quantum spin chain to find partial differential equations determining all the $m$-point correlation functions in the continuum (or thermodynamic) limit. These partial differential equations, the order of which is determined by scaling of the non-equilibrium partition function, are readily solved if we also know the boundary conditions. In this way we can avoid resorting to representation theory of the matrix product algebra. We apply our methods to study the distributions, or moments, of the magnetization and the spin current observables in boundary driven open XXZ spin chains, as well as some connected correlation functions. Interestingly, we find that the transverse connected correlation functions are thermodynamically non-decaying and long-range at the isotropic point $\Delta=1$.
\end{abstract}
\maketitle

\section{Introduction}

Matrix product ansatz (MPA) for steady states has a long history in non-equilibrium physics. MPA was first used to find the non-equilibrium steady states of a 1D asymmetric exclusion model analytically \cite{Derrida} and were also extended to other classical driven diffusive systems \cite{solvers}.  Later MPA has been also applied to steady states of open quantum spin systems, which are described as fixed points of Lindblad master equations \cite{Petruccione}. In these models the spin systems are coupled to two Markovian baths, which drive the system out-of-equlibrium. These types of setups have attracted lots of attention lately in the context of transport theory (see, e.g, Refs. \cite{Prosenreview, ProsenZunkovic, ProsenZnidaric, Prosen1, Prosen2, Arenas, Monasterio, Clark, MCK, HsiangHu, Guo}), as their study has become accessible to experiments through various quantum simulations techniques, such as those using cold atoms \cite{Bloch, JakschZoller, Dorner, Hild}. They also find potential application in the context of e.g., quantum computing \cite{Verstraete1, AlbertJiang1, AlbertJiang2}.  

When we are studying the statistical properties of such systems either the continuum or thermodynamic limit is the most interesting. Due to the richness of phenomena and highly non-trivial nature of out-of-equilibrium systems interesting and useful information can be gained from studying not merely the expectation value of physical quantities, but their statistical properties (fluctuations) as well. For instance, one may be interested in e.g., cumulants, correlators, connected correlation functions in various contexts. 

Calculation of the statistical properties can be greatly eased through the use of large deviation theory and the related method of full-counting statistics, by means of which one can in principle calculate the full probability distributions of physical quantities of interest \cite{reviewfl}. These two methods were only very recently applied in full to quantum systems \cite{Garrahan} and even more recently to both non-interacting (see e.g., \cite{Saito, Znidaric1, KM, Ates}) and interacting \cite{BucaProsen1, Znidaric2, Znidaric3} many body systems. Using these methods open quantum systems can be also seen to exhibit interesting properties near phase transitions \cite{Hickey, Pigeon, ManzanoHurtado, Flindt} and one may also extend them to study closed systems \cite{ChuchemCohen1, ChuchemCohen2}. 

However there are only a few known exact analytical results for the full counting statistics of non-interacting many-body quantum systems \cite{Znidaric1, KM, Znidaric3}, and likewise, there is only a single analytical result for an interacting case of the XXZ spin chain \cite{BucaProsen1}. The latter was achieved only perturbatively in system-bath coupling. In fact,  analytically studying interacting many-body quantum systems under the open quantum system framework seemed like a formidable task, but became feasible after two recent results \cite{Prosen1, Prosen2} for the open XXZ spin chain, later understood through the underlying quantum integrability of the system \cite{PKS, IlievskiProsenLS, IlievskiZunkovic} (for a review see \cite{Prosenreview}). 

Throughout our work we shall call the matrices constituting the matrix product steady state as {\em auxiliary space operators} (ASO). In the aforementioned solutions the ASOs fulfil certain algebraic relations, which we will call the \emph{matrix product algebra}. To compute observables one must usually employ an appropriate representation of the algebra.  We will show however that what one requires, in principle, are only the defining relations of the algebra, which in the continuum limit (or thermodynamic limit) lead to partial differential equations (and of course the corresponding boundary conditions we need to solve these equations). What we study in this article is essentially a simple generalization of the procedure employed in \cite{Prosen2} to calculate the magnetization profiles and 2-point connected longitudinal spin correlation function (the same method was also later used in Ref. \cite{PKS2} to compute the profiles and currents -- but not the correlations -- for more general, so-called twisted boundary conditions). 

Using this method we then find explicit expressions for several $m$-point connected correlation functions and spin current fluctuations for the boundary driven open XXZ spin chain \cite{Prosen2, Prosenreview}. Interestingly, for the critical $\Delta=1$ case the transverse connected correlation functions are non-decaying and thus exhibit genuine long-range order, similar to what has been previously observed numerically in a related case \cite{longrange}. We also study the probability distribution of total magnetization and the moments of the spin current operator. 

Note that we compute these correlators for the non-equilibrium steady state, which can be contrasted with other dynamical studies of both open quantum systems \cite{Bojan1} and systems undergoing quantum quenches \cite{Fagotti}. We should also note that Verstraete and Cirac \cite{VerstraeteCirac} introduced continuous matrix product states (cMPS) for quantum fields. Our approach is not related to this. Instead of constructing matrix product states for quantum fields, which are continuum limits of lattice theories, we will take a discrete matrix product and study the continuum limit. 

In this paper we discuss a general procedure for computing the continuum limit of a steady state (assumed to be given in the form of a matrix product state). A key step in taking this continuum limit is a perturbative expansion in lattice spacing, the validity of which is not known. The second result, is the computation of connected correlators and fluctuations of current in the steady state for the open maximally driven XXZ spin chain in the continuum limit. Using a known discrete solution for this model \cite{Prosen2} we can check the validity of our method. 

More specifically, in Sec.~\ref{sec:mpss} we review the properties of matrix product steady states. In Sec.~\ref{sec:cont} we  outline our method for computing the continuum limit of the steady state equation and steady state (under certain assumption discussed). Later, in Sec.~\ref{sec:xxz} using this method (and also aided with the known solution for the discrete steady state \cite{Prosen2}) we compute the correlation and connected correlation functions for the steady state of the aforementioned open maximally driven XXZ spin chain, focusing mostly on the non-trivial isotropic XXX case. 
Afterwards, in Secs.~\ref{sec:def} and \ref{sec:fluc} we study the fluctuations of spin current and total magnetization in the steady state of the open maximally driven XXZ spin chain, aided by our previous computation of the connected correlation functions. 

\section{Matrix product steady states}
\label{sec:mpss}
We will be interested in non-equilibrium steady states (NESS) $\rho_\infty$ of one-dimensional spin-$1/2$ systems (quantum spin chains). Let the system have $n$ sites described by a $2^n$-dimensional Hilbert space ${\cal H}$ on which act operators constructed from the Pauli matrices, $\sigma^{\pm}_j$, $\sigma^{\z}_j, \sigma^{0}_j := \one $, where $j=1, \ldots, n$ labels the site position.
Let the dynamics of the system, described by the density matrix $\rho(t)$, be determined by a quantum Liouville equation,
\begin{equation}
\frac{\dd}{\dd t}\rho(t)=\LL\rho(t), \label{eqn:0}
\end{equation}
where $\LL$ can be understood as a superoperator acting on the space of operators ${\cal B}(\cal H)$, spanned by the Pauli matrices. The space ${\cal B}({\cal H})$ may also be considered as a Hilbert space itself if one defines an inner product in the Hilbert-Schmidt sense, i.e., $A, B \in {\cal B}({\cal H})$, $(A,B)=\tr A^\dagger B$.
From Eq. \eqref{eqn:0} the defining equation for the NESS is, 
\begin{equation}
\LL\rho_\infty= 0, \label{steady}
\end{equation}
The key assumption we will use is that the NESS is given in the form of a homogenous MPA, 
\begin{equation}
\bb{\rho_\infty}=\frac{S_n}{\tr S_n}\label{eqn:2}
\end{equation}
where, 
\begin{equation}
S_n=\bra{\La}\bb{L}^{\otimes n}\ket{\Ra},\label{eqn:1}
\end{equation}
such that, 
\begin{equation}
S_n=
\bra{\La} \begin{pmatrix}
\bb{O}_{1} & \bb{O}_{-} \cr
\bb{O}_{+} & \bb{O}_{2}
\end{pmatrix} ^{\otimes n} \ket{\Ra},
\label{eqn:3}
\end{equation}
where $\bb{O}_j \in \End ({\frak{S}}_{s})$ are called the auxiliary space operators (ASO)\footnote{These operators were sometimes also called \emph{vertex operators} in Refs.~\cite{Prosenreview, Prosen1, Prosen2}, but they have no relation to the more standard concept of \emph{vertex operators} (see e.g., \cite{Daviesvertex}).} acting over the vector space ${\frak{S}}_{s}$, which is also the space on which the representation of the symmetry algebra of the model acts. Importantly, these operators  satisfy some algebraic relations, which are assumed to be known. The states $\ket{\Ra},\ket{\La} \in {\frak{S}}_{s}$ are referred to as the boundary vectors. We will also define four important operators, $\bb{O}_0$, $\bb{O}_{\rm z}$, $\bb{O}_\x$, and $\bb{O}_\y$,
\begin{align}
\bb{O}_0 &:=\bb{O}_{1}+\bb{O}_{2}, \qquad \bb{O}_{\rm z}:=\bb{O}_{1}-\bb{O}_{2}, \label{vert} \\
\bb{O}_\x &:=\bb{O_+ + O_-} \qquad \bb{O}_\y :=-\ii (\bb{O_+ - O_-}) \label{vertxy}
\end{align}
One usually assumes that the representation (used to calculate the steady state explicitly) of the algebra satisfied by the auxiliary space operators is known. We will not do so here, but will first illustrate the approach one takes if it is known.
Observables can be calculated from the MPA in the following simple way, provided one knows the representation of the ASO algebra. 

Define a general, not necessarily local, operator $B_{\alpha_1, \ldots \alpha_n}= \sigma_1^{\alpha_1}\sigma_2^{\alpha_2} \ldots \sigma_n^{\alpha_n}$, where $1 \ldots n$ label the sites and $\alpha_j \in \{ \x, \y, \z, 0 \}$ denote the components of the corresponding Pauli matrices \cite{Prosenreview, IlievskiPhD}. Its expectation value in the steady state is given by (from Eq. \eqref{eqn:2}),
\begin{equation}
\ave{B_{\alpha_1, \ldots \alpha_n}}=\tr \left( \rho_\infty B_{\alpha_1, \ldots \alpha_n}\right)= \frac{ \tr ( S_n B_{\alpha_1, \ldots \alpha_n})} {\tr S_n}. \label{trcordef}
\end{equation}
An object which will be of central importance is the so-called non-equilibrium partition function, ${\cal{Z}}_{n}$,
\begin{equation}
{\cal{Z}}_{n}=\tr S_n \label{pf}
\end{equation}
It is related to currents flowing through the system in the NESS for a wide variety of exactly solvable 1D systems, including both classical processes \cite{solvers} and various out-of-equilibrium quantum spin chains \cite{Prosenreview}. 

Using the MPA form in Eq. \eqref{eqn:3} Eq. \eqref{trcordef} can be written as, 
\begin{equation}
\ave{B_{\alpha_1, \ldots \alpha_n}}= \frac{\bra{\La}\left[ \tr_{\rm{p}}\left( \begin{pmatrix}
\bb{O}_{1} & \bb{O}_{-} \cr
\bb{O}_{+} & \bb{O}_{2}
\end{pmatrix} ^{\otimes n}  B_{\alpha_1, \ldots \alpha_n}\right)\right] \ket{\Ra}} {\tr S_n},
\end{equation}
where we have used the fact that the trace $\tr:=\tr_{\rm{p}}$ is taken only over the physical $2^n$-dimensional Hilbert space ${\cal H}$ and \emph{not} over the auxiliary space ${\frak{S}}_{s}$ by definition (Eq. \eqref{trcordef}). 
Then it is merely a matter of simple matrix multiplication (in the physical space) and using repeatedly the property of the trace that $\tr(A \otimes B)=\tr(A) \tr(B)$ (together with definitions Eqs. \eqref{vert}, \eqref{vertxy}) to find that, 
\begin{equation}
\ave{B_{\alpha_1, \ldots \alpha_n}}= \frac{\bra{\La} \bb{O}_{\alpha_1} \bb{O}_{\alpha_2} \ldots \bb{O}_{\alpha_n} \ket{\Ra}} {\bra{\La} \bb{O}_0^n \ket{\Ra}}, \label{exp}
\end{equation}
where $\alpha_j \in \{ \x,\y, \z, 0 \} $. Note that this also gives us that the non-equilibrium partition function \eqref{pf} can be written as,
\begin{equation}
{\cal{Z}}_{n}=\tr S_n=\bra{\La}{\bb{O}}^n_0\ket{\Ra}.
\end{equation}
 We can thus define a mapping from expectation values of observables to their corresponding auxiliary space operators, e.g., 
\begin{equation}
\ave{\sigma^{\z}_j \sigma^\z_{k}}=\frac{\bra{\La} \bb{O}_{0}^{j-1}\bb{O}_{\rm z} \bb{O}_{0}^{k-j-1} \bb{O}_{\rm z} \bb{O}_{0}^{n-k} \ket{\Ra}} {\bra{\La} \bb{O}_0^n \ket{\Ra}}.
\end{equation}

In order to actually calculate the expectation values of operators in the NESS we also need know the representation of the ASOs $\bb{O}_j$. For interacting systems these representation are generically infinite dimensional. Even though the representations are near-diagonal in the integrable (in the sense of Ref. \cite{Prosenreview}) cases and thus allow for efficient computation, calculating the expectation values when the operators are not ultra-local (acting only on a single site) or for very large systems is impossible. Using the approach discussed in the next section we show that we can bypass this difficulty (at least up to a multiplicative prefactor) by employing only the asymptotic ($n \rightarrow \infty$) form of the non-equilibrium partition function ${\cal{Z}}_{n}$, together with the algebra satisfied by the ASOs, to calculate all the $m$-point correlators (and equivalently the entire steady state of the system) in the continuum limit without resorting to an explicit representation of the ASOs.

\section{Continuum limit of the NESS} \label{sec:cont}

Our procedure is similar to that used for one and two-point functions in \cite{Prosen2} and later in \cite{PKS2}. Motivated by these known examples we will consider only ASOs which satisfy at most cubic algebraic relations of the form,
\begin{align}
&k^{\alpha}_{3,1} \bb{O}_\alpha \bb{O}_0 \bb{O}_0+k^{\alpha}_{3,2} \bb{O}_0 \bb{O}_\alpha \bb{O}_0+k^{\alpha}_{3,3} \bb{O}_0 \bb{O}_0 \bb{O}_\alpha \nonumber \\
+&k^{\alpha}_{2,1} \bb{O}_\alpha \bb{O}_0+  k^{\alpha}_{2,2} \bb{O}_0 \bb{O}_\alpha+k^{\alpha}_{1,1}  \bb{O}_\alpha =0, \label{alg}
\end{align}
where $\alpha \in \{\x,\y,\z\}$ for some constants $k^{\alpha}_{i,j} $. In principle our approach can be generalized to other cases as well, though we will not discuss this here.
Let us now introduce a {\em lattice spacing} $a=1/n$, such that the total length of the system is unity, and so the continuum limit $a\to 0$ corresponds to the thermodynamic limit 
 $n \rightarrow \infty$. 

We wish to find a set of partial differential equations for a $m$-point correlator, $C^{\alpha_1,\alpha_2,\ldots,\alpha_m}_{j_1, j_2,\ldots, j_m}=\ave{\sigma^{\alpha_1}_{j_1} \sigma^{\alpha_2}_{j_2} \ldots \sigma^{\alpha_m}_{j_m}} $, where $\alpha_j \in \{ \x, \y, \z \}$ for each of the $m$ operator coordinates. 

We will first find them for $\alpha_1$ ($j_1$) by multiplying  Eq. \eqref{alg} for $\alpha=\alpha_1$, by $\bb{O}_0^{j_2-j_1-1}\bb{O}_{\alpha_2}\bb{O}_0^{j_3-j_2-1}\bb{O}_{\alpha_3}\ldots \bb{O}_{\alpha_m} \bb{O}_0^{n-j_m} \ket{\Ra}$
from the right and by $\bra{\La} \bb{O}_0^{j_1-1}$ from the left and divide it by the non-equilibrium partition function ${\cal{Z}}_{n}=\bra{\La}\bb{O}_0^n \ket{\Ra} $. We then use Eq. \eqref{exp} to find similarly to \cite{PKS2, Prosenreview},
\begin{align}
&k^{\alpha_1}_{3,1} C_{j_1,j_2+3, \ldots, j_m+3}^{\alpha_1,\ldots,\alpha_m;n+1}+k^{\alpha_1}_{3,2}  C_{j_1+1,j_2+3, \ldots, j_m+3}^{\alpha_1,\ldots,\alpha_m;n+1}+k^{\alpha_1}_{3,3}  C_{j_1+2,j_2+3, \ldots, j_m+3}^{\alpha_1,\ldots,\alpha_m;n+1} \nonumber \\
+&(k^{\alpha_1}_{2,1} C_{j_1,j_2+2, \ldots, j_m+2}^{\alpha_1,\ldots,\alpha_m;n}+  k^{\alpha_1}_{2,2} C_{j_1+1,j_2+2, \ldots, j_m+2}^{\alpha_1,\ldots,\alpha_m;n})\frac{{\cal{Z}}_{n-1}}{{\cal{Z}}_{n}}+k^{\alpha_1}_{1,1}  C_{j_1,j_2+1, \ldots, j_m+1}^{\alpha_1,\ldots,\alpha_m;n-1} \frac{{\cal{Z}}_{n-2}}{{\cal{Z}}_{n}}=0, \label{cor}
\end{align}
where the superscript $n$ over $C_{j_1 \ldots}^{\alpha_1,\ldots,\alpha_m;n}$ denotes that the this correlator is computed for system size $n$.
Define $x_k=j_k/n$, $x_k+a:=(j_k+1)/n$, with inter-site (lattice) spacing denoted as $a:=1/n$, and likewise $C^{\alpha_1,\ldots \alpha_m}(x_1 \ldots x_m):= C^{\alpha_1,\ldots \alpha_m}_{j_1 \ldots j_m }$. Note that like in \cite{Prosenreview, Prosen2, PKS2} one may instead equivalently define $x_k=(j_k-1)/(n-1)$ and $a=1/(n-1)$ as we will do in the next section. Assume that we can expand for large $n$ as,
\be
\frac{{\cal{Z}}_{n-1}}{{\cal{Z}}_{n}}=\sum_{m=0}^{\infty} z^{(m)}n^{-m}, \label{partfunexpans}
\ee 
where first few coefficients $z^{(0)}, z^{(1)}$, etc, may be vanishing. As we will see later, this type of expansion can be performed for the open maximally driven XXZ spin chain at $\Delta \leq 1$ and for some other models such as open $SU(N)$-symmetric quantum gases \cite{IlievskiZunkovic}, XXZ spin chains with twisted boundary driving (also for $\Delta \leq 1$) \cite{PKS, PKS2}, the open spin-1 Lai-Sutherland chain \cite{IlievskiProsenLS} and some other cases discussed in the review article \cite{Prosenreview}. 
We then take the continuum limit $n \rightarrow \infty$ by expanding in $1/n$\footnote{Note that we have suppressed the superscript $\alpha_1 \ldots \alpha_m$ in $ C^{(k)}(x_1, \ldots x_m)$; when we do this we refer to a \emph{general} $m$-point correlator, i.e., $C(x_1, \ldots x_m)=C^{\alpha_1,\ldots, \alpha_m}(x_1, \ldots x_m)$.  },
\begin{align}
 C_{j_1 \ldots}^{\alpha_1,\ldots,\alpha_m;n}&=C^{(0)}(x_1, \ldots, x_m)+\frac{C^{(1)}(x_1, \ldots, x_m)}{n}+{\cal O}(\frac{1}{n^2}), \qquad x_k=\frac{j_k}{n} \label{expandcor1}
\end{align}
and performing Taylor series expansion\footnote{We assume that we can perform this expansion. The points when the indices coincide can introduce extra boundary conditions and even cause our expansion to fail when these indices are close to each other in the continuum limit. For the open maximally driven XXZ case we study later we have the added benefit of having a discrete solution for the NESS which was previously found to check our results \cite{Prosen2}.} in $a=1/ n$ around 0 to find an infinite system of partial differential equations up to arbitrary order in $1/n$ (where, for brevity, $k^{\alpha_1}_3:=k^{\alpha_1}_{3,1} +k^{\alpha_1}_{3,2} +k^{\alpha_1}_{3,3}$, $k^{\alpha_1}_2:=k^{\alpha_1}_{2,1}+k^{\alpha_1}_{2,2}$ and $C^{(k)}(x_1, \ldots, x_m):=C^{(k)}(\vec{x})$),
\begin{align}
&(z^{(0)}k^{\alpha_1}_2 + k^{\alpha_1}_3)C^{(0)}(\vec{x})=0, \nonumber \\
&z^{(1)}k^{\alpha_1}_2  C^{(0)}(\vec{x})+ \nonumber \\
&(z^{(0)}k^{\alpha_1}_2+k^{\alpha_1}_3) C^{(1)}(\vec{x})+(k^{\alpha_1}_{3,3}- k^{\alpha_1}_{3,1}+[( k^{\alpha_1}_{2})x_1 -k^{\alpha_1}_{2,1}])  \partial_{x_1} C^{(0)}(\vec{x}))+\nonumber \\ 
&z^{(0)}(k^{\alpha_1}_{2} \sum_{j \neq 1}(x_j-1)\partial_{x_j} C^{(0)}(\vec{x})=0, \nonumber\\
& k^{\alpha_1}_{2} (z^{(2)} C^{(0)}(\vec{x})+z^{(1)} C^{(1)}(\vec{x}))+\nonumber\\
&(k^{\alpha_1}_3+z^{(0)}k^{\alpha_1}_3) C^{(2)}(\vec{x})+k^{\alpha_1}_{2}\sum_{j \neq 1} \{ (x_j-1)(z^{(0)}\partial_{x_j} C^{(1)}(\vec{x})+z^{(1)}\partial_{x_j} C^{(0)}(\vec{x}))+\nonumber\\
&\frac{1}{2}k^{\alpha_1}_2 (x_j-1)^2 z^{(0)}\partial_{x_j}^2 C^{(0)}(\vec{x}) \}+ (k^{\alpha_1}_{2,1} ( x_1-1) + k^{\alpha_1}_{2,2} x_1)z^{(0)}\partial_{x_1} C^{(0)}(\vec{x})+\nonumber\\
&(k^{\alpha_1}_{3,3} - k^{\alpha_1}_{3,1} + ( k^{\alpha_1}_{2} x-k^{\alpha_1}_{2,1}) z^{(0)})\partial_{x_1} C^{(1)}(\vec{x})+\nonumber\\
&\sum_{j \neq 1}(k^{\alpha_1}_{2,1} (x_1-1) + k^{\alpha_1}_{2,2} x_1) (x_j -1) z^{(0)}\partial_{x_1}\partial_{x_j} C^{(0)}(\vec{x})+ \nonumber\\
&\frac{1}{2} [k^{\alpha_1}_{3,1} + k^{\alpha_1}_{3,3} + (k^{\alpha_1}_{2,1} (1- 2 x_1) + k^{\alpha_1}_{2} x_1^2) z^{(0)}]\partial_{x_1}^2 C^{(0)}(\vec{x})=0\nonumber\\
 &\vdots \label{set}
\end{align}
We then continue by deriving the equations for $j_2$ by first multiplying Eq. \eqref{alg} by $\bra{\La} \bb{O}_0^{j1-1}\bb{O}_{\alpha_1}\bb{O}_0^{j_2-j_1-1} $ from the left, etc. and likewise for all $\alpha_j$. 
We are finally left with a set of coupled partial differential equations for every order in $1/n$. These are in general complicated for arbitrary orders, but if we only focus on the leading order $(1/n)^0$ they are generically quite simple. The leading order will be determined by the first non-zero equation in the set Eq. \eqref{set}. For instance if $z^{(0)}k^{\alpha_1}_2 + k^{\alpha_1}_3 \neq 0$ the leading order is given by $C^{(0)}(\vec{x})=0$. 

Note that everything, except the boundary conditions, is fully determined by the algebraic relations Eq. \eqref{alg} and the asymptotic scaling $n \rightarrow \infty$ of the non-equilibrium partition function ${\cal{Z}}_n$ Eq. \eqref{pf}. 

We did not consider the representation of the ASOs at all. One may object that the representation is relevant when one wishes to find the boundary conditions to solve these partial differential equations and that it comes into play via the boundary vectors $\ket{\Ra}$ and $\ket{\La}$, which we used when deriving Eq. \eqref{cor}. However, in the leading order at least this can be circumvented for a quite general set of Liouvillians  Eq. \eqref{eqn:0} which define the non-equilibrium steady state Eq. \eqref{steady},
\begin{equation}
\LL\rho_\infty=(\LL_0 + \LL_1 + \LL_n)\rho_\infty= 0 \label{steadyagain}
\end{equation} 
 where $\LL_0$ acts in general in the bulk of the system and $\LL_1$ and $\LL_n$ act only ultralocally (on the boundary sites $1$ and $n$, respectively). These are the types of Liouvillians one most often encounters when studying non-equlibrium matrix product steady states.
 
As mentioned previously, we work with one-dimensional spin-1/2 systems and thus we take that $\LL_0$ is associated with a model with finite lattice spacing $a$. For instance, in the case of the Lindblad master equation (which we will study in more detail in the next section) $\LL_0 \rho=-\ii [H,\rho]$, where $H$ is the Hamiltonian of the system. 

Assume that $\LL_0$ can be written in terms of local two-site interaction operators, $\LL_0=\sum_j \sum_d {\lld}^d_{j}{\lld}^d_{j+1}$, where $j,j+1$ denotes the sites on which the operator acts. We then perform the continuum limit as before by first setting $a=1/n$, $x=j/n$, $x+a=(j+1)/n$. Namely, 
\be
\lld^d_{j} \to \lld(x=\frac{j}{n}) \qquad \lld^d_{j+1} \to \lld(x+a=\frac{j+1}{n}) \label{conttrans}
\ee
Formally then, when taking the continuum limit, $a=1/n \rightarrow 0$ we may expand for small $a$, $\LL_0(a)=\LL_0^{(0)}+ {\cal{O}}(a)$, where the $a$ in $\LL_0(a)$ denotes that we are now dealing with an operator which depends on lattice spacing $a$ after we used Eq. \eqref{conttrans}.

 Likewise we may formally take the continuum limit of the NESS by first writing out in the operator basis, 
 \be
 \rho_\infty=N\left(\one+\sum_k \sum_\alpha \ave{\sigma^\alpha_k}\sigma^\alpha_k+\sum_{k \neq m} \sum_{\alpha, \beta} \ave{\sigma^\alpha_k \sigma^\beta_m}\sigma^\alpha_k\sigma^\beta_m \ldots \right), \label{thocor}
 \ee
 where $N$ is a normalization coefficient (such that $\tr \rho_\infty$=1), and then taking the same continuum limit $\rho_\infty(a)=\rho^{(0)}+\rho^{(1)} a +{\cal{O}}(a^2)$. We do this in the following manner. First rewrite Eq. \eqref{thocor} as discussed, 
  \begin{align}
 &\rho_\infty=N\left(\one+\sum_{k} \sum_\alpha \ave{\sigma^\alpha(x=k a)}\sigma^\alpha(x=ka) \right. \nonumber \\
 &\left.+\sum_{k \neq m } \sum_{\alpha, \beta} \ave{\sigma^\alpha(x_1=ka) \sigma^\beta(x_2=ma)}\sigma^\alpha(x_1=ka)\sigma^\beta(x_2=ma) \ldots \right). \label{thocor1}
 \end{align}
 We then formally expand the correlation functions in Eq. \eqref{thocor1} for small $a=1/n$, $ \ave{\sigma^\alpha(x=ka}= \ave{\sigma^\alpha(x)}^{(0)}+ \ave{\sigma^\alpha(x)}^{(1)} a+{\cal O}(a^2), \ave{\sigma^\alpha(x_1=ka) \sigma^\beta(x_2=ma)}=\ave{\sigma^\alpha(x_1) \sigma^\beta(x_2)}^{(0)}  + \ave{\sigma^\alpha(x_1) \sigma^\beta(x_2)}^{(1)} a +{\cal O}(a^2), \ldots$ Plugging this back into Eq. \eqref{thocor1} and gathering terms in $a$ we identify,
 \begin{align}
&\rho^{(0)}=N\left(\one+ \lim_{a=\frac{1}{n} \to 0}\sum_k\sum_\alpha \ave{\sigma^\alpha(x=ka)}^{(0)}\sigma^\alpha(x=ka)\right. \nonumber \\
&\left.+\lim_{a=\frac{1}{n} \to 0} \sum_{k \neq m} \sum_{\alpha, \beta} \ave{\sigma^\alpha(x_1=ka) \sigma^\beta(x_2=ma)}^{(0)}\sigma^\alpha(x_1=ka)\sigma^\beta(x_2=ma) \ldots \right) \label{steadystateexpanded1} \\
&\rho^{(1)}=N\left(\lim_{a=\frac{1}{n} \to 0}\sum_k\sum_\alpha \ave{\sigma^\alpha(x=ka)}^{(1)}\sigma^\alpha(x=ka)\right.\nonumber \\
&\left.+\lim_{a=\frac{1}{n} \to 0} \sum_{k \neq m} \sum_{\alpha, \beta} \ave{\sigma^\alpha(x_1=ka) \sigma^\beta(x_2=ma)}^{(1)}\sigma^\alpha(x_1=ka)\sigma^\beta(x_2=ma) \ldots \right). \label{steadystateexpanded2}
\end{align}
Let us pause to make a few comments. When the difference between $x_1$ and $x_2$, etc. is of the order of lattice spacing $a$ the expansion may be ill-defined. In fact, this will turn out to be the case for the $\Delta<1$ case latter studied. When we know the discrete solution for the NESS, which will be the case when we will later study the open XXZ spin chain, we can use this solution to check our results for the continuum limit. Otherwise, one may have to simply assume that the continuum limit can be taken. Furthermore, the normalization coefficient $N$ depends on $a=1/n$. However, it can be seen to cancel in the equation for the NESS \eqref{steadyagain} and thus does not influence the physical results. 

 We also assume that $\LL_1$ and $\LL_n$ do not depend on $a$. We then have in the leading order $a^0$,
\begin{equation}
(\LL_1 + \LL_n+\LL_0^{(0)})\rho^{(0)}= 0, \label{zero}
\end{equation} 
where $\rho^{(0)}$ is essentially almost equivalent to knowledge of all the correlators in the continuum limit. 

Note that this continuum limit is the same as the one we took when calculating the differential equations for the correlators Eq. \eqref{set}. In other words, if we already know the discrete solution for the NESS $\rho_\infty$ taking the continuum limit for the correlators as we did when finding Eq. \eqref{set} also gives the solution perturbatively in lattice spacing $a$. Since $\LL_1$ and $\LL_n$ are ultralocal (assumed to be acting on one site each) solving Eq. \eqref{zero} and thus obtaining the boundary conditions needed to solve the leading order of the set of partial differential equations in Eq. \eqref{set} is simple.  

It is important to note that even though we may circumvent the issue of not knowing the boundary vectors $\ket{\Ra}$ and $\ket{\La}$ and the representation of the ASOs using the above discussed approach, it is not necessary to do so. In case of the already solved problem of the steady state of the open non-equilibrium boundary driven XXZ spin chain \cite{Prosen2} in terms of an MPA the boundary vectors and the representation of the ASOs are known. One can then simply use this to find the appropriate boundary conditions when solving the partial differential equations \eqref{set} in a manner similar to what was done in \cite{Prosen2, PKS2} for a less general set of correlators. 

We will use Eq. \eqref{zero} in the next section to find the boundary conditions for the partial differential equations determining the NESS of a boundary driven open XXX spin chain.

We will now turn to an example of a previously solved MPA steady state of an open non-equilibrium boundary driven XXZ spin chain and compute the $m$-point correlators in this case.

\section{The $m$-point correlators of the maximally boundary driven XXZ spin chain} \label{sec:xxz}

The Lindblad master equation is a useful tool for describing out-of-equilibrium physics \cite{Lindblad}. It can represent both driving and decoherence by a set of infinite baths coupled to a system under the Born-Markov and rotating wave approximations \cite{Petruccione}. It also has an important property of being the most general form of a time-local Markovian quantum master equation which is both completely positive and trace preserving. The Lindblad master equation is,
\begin{align}
\frac{\dd}{\dd t}\rho(t)&=\LL \rho(t)=-\ii [H,\rho(t)]+\DD(\rho(t)),\nonumber \\
\DD(\rho(t))&:=\sum_{k}\Gamma_{k}\left(L_{k}\rho(t)L^{\dagger}_{k}-\half\left\{L^{\dagger}_{k}L_{k},\rho(t)\right\}\right),
\label{lindblad}
\end{align}
where we will take $H$ to be the XXZ spin chain Hamiltonian,
\begin{eqnarray}
H=\sum_{j=1}^{n}2(\sigma^{+}_{j}\sigma^{-}_{j+1}+\sigma^{-}_{j}\sigma^{+}_{j+1})+
\Delta \sigma^{\rm{z}}_{j}\sigma^{\rm{z}}_{j+1}, \label{ham}
\end{eqnarray}
and Lindblad operators acting only on the boundary sites $1$ and $n$,
\begin{equation}
L_{1}=\sqrt{\varepsilon}\sigma^{+}_{1},\qquad L_{n}=\sqrt{\varepsilon}\sigma^{-}_{n}.
\label{maxdriv}
\end{equation}
They represent maximum driving, that is the left bath decoherently only injects magnetization into the system and the right one only takes it out. 
The equation for the NESS is exactly solvable (where we define the superoperator $(\!\ad H)\rho \equiv [H,\rho]$),
\begin{equation}
\LL \rho_{\infty}=-\ii \ad H\rho_{\infty}+\DD(\rho_{\infty})=0, \label{steadystate}
\end{equation}
due to the underlying integrability structure of the XXZ spin chain. The solution was found in \cite{Prosen2} (see \cite{Prosenreview} for a more comprehensive overview). 

\subsection{The isotropic point $\Delta=1$}

At the isotropic point $\Delta=1$, it is known that the ASOs satisfy the following cubic algebraic relations \cite{Prosenreview, PKS2, IlievskiPhD}, 
\begin{equation}
[\bb{O}_0,[\bb{O}_0,\bb{O}_\alpha]]+2\{\bb{O}_0,\bb{O}_\alpha\}-8p^2 \bb{O}_\alpha=0, \qquad \alpha=\x,\y,\z, \qquad p=\frac{4 \ii} {\varepsilon}. \label{spin2}
\end{equation}
It is also known that for $\varepsilon \gg 1/n$ the partition function scales as \cite{Prosenreview},
\begin{equation}
\frac{{\cal{Z}}_{n-1}}{{\cal{Z}}_{n}}=\frac{\pi^2 }{4 n^2}+ {\cal{O}}(n^{-3})). 
\end{equation}
Using the method discussed in the previous section we immediately arrive at a set of decoupled second order partial differential equations for the $m$-point correlators in the continuum limit\footnote{Note that it turns out due to the cubic algebra of this problem Eq. \eqref{spin2} that the leading order terms after performing the expansion via Eq. \eqref{set} do not depend on $\varepsilon$ and $p$.},
\begin{equation}
\frac{\partial^2}{\partial x_k ^2} C^{(0)}(\vec{x}) = -\pi^2 C^{(0)}(\vec{x}) \qquad k=1, \ldots, m, \label{leadingorderp}
\end{equation}
where we emphasize again to avoid confusion, that $C^{(0)}(\vec{x})$ denotes the leading order in $1/n$ of a general $m$-point correlator $C(\vec{x})$,
\be
C(\vec{x}):=C(\sigma^{\alpha_1}_{j_1}\sigma^{\alpha_2}_{j_2} \ldots \sigma^{\alpha_m}_{j_m}):=\ave{\sigma^{\alpha_1}(x_1)\sigma^{\alpha_2}(x_2) \ldots \sigma^{\alpha_m}(x_m)}, \qquad x_k=\frac{j_k}{n},
\ee
i.e,
\be
C(\vec{x})=\sum_{k=0}^\infty C^{(k)}(\vec{x})\left(\frac{1}{n}\right)^k
\ee
Now we will show how one can obtain the boundary conditions as outlined briefly in the previous section, Sec.~\ref{sec:cont}.  First we take $x=j/n=j a$ and set $\sigma^\alpha_j \to \sigma^\alpha(x) $, $\sigma^\alpha_{j+1} \to \sigma^\alpha(x+a) $ in the Hamiltonian \eqref{ham} (for $\Delta=1$), while keeping the length fixed $n a=1$, and then expand in lattice spacing $a$.  To do this first look at a pair local densities $h_{j,j+1}=\sum_{\alpha=\x,\y,\z}\sigma^\alpha_j\sigma^\alpha_{j+1}$ in the discrete Hamiltonian \eqref{ham} (noting that $H=\sum_{j=1}^{n-1}h_{j,j+1}$), which in the continuum limit go as, 
\be
h_{j,j+1}+h_{j-1,j} \to h(x,x+a)+h(x-a,x)=\sum_{\alpha=\x,\y,\z}\sigma^\alpha(x-a)\sigma^\alpha(x)+\sigma^\alpha(x)\sigma^\alpha(x+a).
\ee
Expanding in $a$, we find,
\be
 h(x,x+a)+h(x-a,x)=\sum_{\alpha=x,y,z} 2\sigma^\alpha(x)\sigma^\alpha(x) +2\sigma^\alpha(x) \frac{\partial^2 \sigma^\alpha(x)}{\partial^2 x} a^2+{\cal{O}}(a^3),
\ee
i.e., the term of order $a$ containing the first derivative cancels and the first non-trivial term is the one with the second order derivative. It may be interesting to note that this cancellation of the terms with the first derivatives mimics the one we got when deriving Eq. \eqref{leadingorderp}.
Now using the (Riemann) sum definition of the integral and the basic algebra of the Pauli matrices we arrive that in the leading orders of $a$ the Hamiltonian is,
\be
H=\frac{3}{a}-6+a \int_0^1  dx \sum_{\alpha=\x,\y,\z} \left( \sigma^\alpha(x) \frac{\partial^2 \sigma^\alpha(x)}{\partial^2 x} \right)  + {\cal{O}}(a^2). \label{hamexpand1}
\ee
The leading term here in Eq. \eqref{hamexpand1} is divergent as $a \to 0$. However, this is not much of a problem as this term commutes with any operator in $ \ad H $ in the steady state equation \eqref{steadystate} and thus has no influence on the final result for the NESS. Likewise the term of order $a^0$ commutes with any operator and does not influence $\ad H$. 

Therefore, the first non-trivial term in Eq. \eqref{hamexpand1} is of order $a$. We note for the interested reader that, after using integration by parts, this term corresponds to what is known as the quantum Landau-Lifshitz model (e.g., \cite{ Minahan,Melikyan}) or the $SU(2)$ quantum continuous Heisenberg magnet \cite{Sklyanin}. 

We can also expand $\rho_\infty$ using Eq. \eqref{thocor} for small $a$.
Then the leading order in $a$ is given by (using Eqs. \eqref{zero}, \eqref{steadystateexpanded1} and Eq. \eqref{steadystate}), 
 \be
 \DD(\rho^{(0)})=0, \label{steadystate0}
 \ee
where $\DD$ is given by Eq. \eqref{lindblad} and $L_1=\sqrt{\varepsilon} \sigma^{+}(0)$, $L_n=\sqrt{\varepsilon} \sigma^{-}(1)$. Note that assuming that $\varepsilon \ne 0$ we can cancel it in Eq. \eqref{steadystate0}. 
Splitting the dissipator into $\DD=\DD_1+\DD_n$, where $\DD_{1,n} \rho=\left(L_{1,n}\rho L^{\dagger}_{1,n}-\half\left\{L^{\dagger}_{1,n}L_{1,n},\rho\right\}\right)$, we observe the following action on the basis operators,
\begin{align}
&\DD_1(\one)=\varepsilon \sigma^\z(0), \qquad &\DD_1(\sigma^\z(0))=-\varepsilon\sigma^\z(0)&, \qquad &\DD_{1}(\sigma^{\x,\y}(0))=-\frac{\varepsilon}{2}\sigma^{\x,\y}(0), \nonumber \\
&\DD_n(\one)=-\varepsilon \sigma^\z(1), \qquad &\DD_n(\sigma^\z(1))=\varepsilon\sigma^\z(1)&, \qquad &\DD_{n}(\sigma^{\x,\y}(1))=-\frac{\varepsilon}{2}\sigma^{\x,\y}(1). \label{disbasis}
\end{align}
Using this Eq. \eqref{disbasis} and requiring a solution to Eq. \eqref{steadystate0} for each of the operators in the basis Eq. \eqref{thocor} we arrive to the following boundary conditions,
\begin{align}
&C^{(0)}(\vec{x}) |_{\sigma^\z(x_k=0)}=\ave{\sigma^{\alpha_1}(x_1) \ldots\sigma^{\alpha_{k-1}}(x_{k-1})\sigma^{\z}(x_k=0)\sigma^{\alpha_{k+1}}(x_{k+1}) \ldots \sigma^{\alpha_m}(x_m)}= \nonumber \\
&\ave{\sigma^{\alpha_1}(x_1) \ldots\sigma^{\alpha_{k-1}}(x_{k-1})\sigma^{\alpha_{k+1}}(x_{k+1}) \ldots \sigma^{\alpha_m}(x_m)}, \label{boundc1} \\
&C^{(0)}(\vec{x}) |_{\sigma^\z(x_k=1)}=\ave{\sigma^{\alpha_1}(x_1) \ldots\sigma^{\alpha_{k-1}}(x_{k-1})\sigma^{\z}(x_k=1)\sigma^{\alpha_{k+1}}(x_{k+1}) \ldots \sigma^{\alpha_m}(x_m)}= \nonumber \\
&-\ave{\sigma^{\alpha_1}(x_1) \ldots\sigma^{\alpha_{k-1}}(x_{k-1})\sigma^{\alpha_{k+1}}(x_{k+1}) \ldots \sigma^{\alpha_m}(x_m)}, \label{boundc2} \\
&C^{(0)}(\vec{x})|_{\sigma^{\x,\y}(x_k=0,1)}=\ave{\sigma^{\alpha_1}(x_1) \ldots\sigma^{\alpha_{k-1}}(x_{k-1})\sigma^{\x,\y}(x_k=0,1)\sigma^{\alpha_{k+1}}(x_{k+1}) \ldots \sigma^{\alpha_m}(x_m)}=0. \label{boundc3} 
\end{align}
The first order equation for the Liouvillian would be given by continuing the expansion of Eq. \eqref{steadystate} in lattice spacing $a=1/n$, using Eq. \eqref{hamexpand1} (for the Hamiltonian) and the explicit forms of the NESS expansion given by Eqs. \eqref{steadystateexpanded1} and \eqref{steadystateexpanded2},
\be
-\ii \ad H^{(1)} \rho^{(0)}+\DD \rho^{(1)}=0.
\ee
However, we do not need this as we already know the algebraic relations of the ASOs defining the steady state (Eq. \eqref{spin2}) which fully determine the leading order $\rho^{(0)}$ via the partial differential equations \eqref{leadingorderp} \emph{up to the boundary conditions for these equations}. 
We have found these boundary conditions for the $m$-point correlators using the preceding discussion and they are given in Eqs. \eqref{boundc1}, \eqref{boundc2} and \eqref{boundc3}, and thus we find,
\begin{equation}
C^{(0)}(\vec{x})=A \prod_{k=1}^m \sin(a_k \frac{\pi}{2}+\pi x_k), \label{cor0}
\end{equation}
where $a_k=0$ if the operator depending on $k-$th coordinate $j_k$ is $\sigma^\x$ or $\sigma^\y$ (i.e., $\alpha_k=\x,\y$) and $a_k=1$ if it is $\sigma^\z$ (or $\alpha_k=\z$) and $A$ is some yet to be determined constant. 

Let us recap: we have used the expansion of the equation for the NESS \eqref{steadystate} in lattice spacing $a=1/n$ and via Eq. \eqref{zero} found the boundary conditions for the NESS of an open XXZ spin chain under boundary driving given by Eq. \eqref{maxdriv}.  It is important to note that this preceding discussion on the perturbation expansion in $a$ (via Eq. \eqref{zero}), which we used \emph{only} to find the boundary conditions for the partial differential equations \eqref{leadingorderp}, is not necessary provided that we also know the action of the representation of the ASOs (which define the MPA for the NESS in Eq. \eqref{eqn:1}), i.e.,  we can use how the ASOs act on the boundary vectors $\ket{\Ra}$($\ket{\La}$) to find the boundary conditions. This is in fact known for the maximally driven open XXZ spin chain \cite{Prosen2} and some of the boundary conditions we found here in Eqs. \eqref{boundc1}, \eqref{boundc2}, \eqref{boundc3} were found in the same article.   

Furthermore, in order to find the scaling factor $A$ we need to employ the boundary conditions given by the appropriate representation of the ASOs \cite{Prosen2}. Doing this for fairly large system sizes we find the following behaviour; if we let $p_x$ denote the number of $\sigma^\x$ and $p_y$ the number of $\sigma^\y$ operators in the correlator then $A=0$ if $p_x$ or $p_y$ are odd, otherwise $A$ is given by the following recurrence relation ($p:=p_x+p_y$) \cite{Sloane},
\[A(p,p_x-p_y) = \left\{
  \begin{array}{lr}
    \frac{(2 p)!}{(m!)^2} & p_x-p_y = 0 \\
 \frac{1}{(m!)^2} (4 A(p-1,p_x-p_y-1) -A(p,p_x-p_y-1)) & p_x-p_y \neq 0   \end{array}
\right.
\]
We have not been able to prove this in general, however exact calculations for up to $n=20$ support this conjecture. 
The initial terms $A(p,0)$ and $A(p+1,0)$ can be calculated using the first line of the recurrence relation and, in turn, can be used to compute $A(p+1,1), A(p+1,2), \ldots$. Also we take $A(0,k):=0$.


In order to calculate with higher precision than $ {\cal{O}}(n^0)$ we will again need to employ the boundary conditions given by the appropriate representation of the ASOs \cite{Prosen2}. These corrections will be important when we want to look at the connected correlation functions. For higher precision in $1/n$ we will likewise need a more precise scaling of the non-equilibrium partition function found in \cite{Prosenreview},
\begin{equation}
\frac{{\cal{Z}}_{n-1}}{{\cal{Z}}_{n}}=\frac{\pi^2 }{4(n-\alpha)^2}(1+ {\cal{O}}(n^{-2})). \label{scale}
\end{equation}
Interestingly, the constant $\alpha$ can be found through self-consistency conditions imposed by the algebra \eqref{spin2} and the boundary conditions, as will be shown later. We find, using Taylor series expansion from Eq. \eqref{set} for Eq. \eqref{spin2}, that the $1/n$ corrections $ C^{(1)}$ are given by the following coupled partial differential equations (omitting $\Delta=1$, setting $\beta=4(1-\alpha)$ and switching to a more compact PDE notation), 
\begin{align}
&C^{(1)}_{x_k x_k} + \pi^2 C^{(1)}=  \label{foe} \\
&\sum_{p<k,j>k}\frac{\pi^2}{2}\left ( (\beta-4) C^{(0)}(\vec{x})+2(1-x_{j}) C^{(0)}_{x_j}+2 x_p  C^{(0)}_{x_p}-(1-2 x_k) C^{(0)}_{x_k} \right )+2C^{(0)}_{x_k x_k}. \nonumber
\end{align}
Likewise we may find the higher order corrections. The PDEs determining them become exceedingly complicated and we omit writing them explicitly. 

Note that these correction terms actually depend on the whether we expand the series in $1/n$ or $1/(n-1)$ (not the definition of $x_k$ directly). The $k$-th order corrections are then of either order in $1/n^k$ or $1/(n-1)^k$. Naturally, the leading orders $1/n^0$ and $1/(n-1)^0$ are equal, but the higher orders are not. This merely reflects the fact that the higher orders give increasingly precise corrections for large, but still finite, system size $n$ and it is merely a matter of convention in what we will expand - they gives equivalent and consistent information about the correlation functions. 

We can now turn to computing the connected correlation functions. They are given by the standard generating function,
\begin{equation}
\widetilde{C}( \sigma_1^{\alpha_1}\sigma_2^{\alpha_2} \ldots \sigma_m^{\alpha_m})= \ave{\sigma_1^{\alpha_1}\sigma_2^{\alpha_2} \ldots \sigma_m^{\alpha_m}}_c:=  \frac{\partial}{\partial z_1}\cdots \frac{\partial}{\partial z_m}\log \tr \rho_{ \infty} (\exp\sum_j z_j \sigma_j^{\alpha_j})\big|_{z_j=0}. \label{corgenfun}
\end{equation}
We will write out the two and three-point connected correlators explicitly, 
\begin{align}
\widetilde{C}(\sigma_{j_1}^{\alpha_{j_1}}\sigma_{j_2}^{\alpha_{j_2}})&=\ave{\sigma^{\alpha_1}_{j_1} \sigma^{\alpha_2}_{j_2}} -\ave{\sigma^{\alpha_1}_{j_1}}\ave{ \sigma^{\alpha_2}_{j_2}}\\ \nonumber
\widetilde{C}(\sigma_{j_1}^{\alpha_{j_1}}\sigma_{j_2}^{\alpha_{j_2}}\sigma_{j_3}^{\alpha_{j_3}})&=\ave{\sigma_{j_1}^{\alpha_{j_1}}\sigma_{j_2}^{\alpha_{j_2}}\sigma_{j_3}^{\alpha_{j_3}}}-\ave{\sigma_{j_1}^{\alpha_{j_1}}}\ave{\sigma_{j_2}^{\alpha_{j_2}}\sigma_{j_3}^{\alpha_{j_3}}}-\ave{\sigma_{j_2}^{\alpha_{j_2}}}\ave{\sigma_{j_3}^{\alpha_{j_3}} \sigma^{\alpha_1}_{j_1}} \\ \nonumber
&-\ave{\sigma^{\alpha_3}_{j_3}}\ave{\sigma^{\alpha_1}_{j_1}\sigma^{\alpha_2}_{j_2}}+2\ave{\sigma^{\alpha_1}_{j_1}}\ave{\sigma^{\alpha_2}_{j_2}}\ave{\sigma^{\alpha_3}_{j_3}}
\end{align}
The expressions for higher connected correlators are quite long so we omit writing them. Now using Eq. \eqref{corgenfun} and the previous results from Eq. \eqref{foe} (as well as the higher order equations obtained from Eq. \eqref{set}) and the boundary conditions given by the matrix representation of the ASOs \cite{Prosen2, Prosenreview} we can find arbitrary $m$-point connected correlators. 

We will show how to obtain some of the lower $m$-point connected correlators. We first note that, like the correlators Eq. \eqref{cor0}, all the connected correlators which contain an odd number of $\sigma^\x$ and $\sigma^\y$ operators are 0. 

First let us then start with the 2-point connected correlators; $\widetilde{C}(\sigma_{j_1}^{\z}\sigma_{j_2}^{\z})$ was already found in \cite{Prosen2, Prosenreview}. It was shown to scale inversely with  the system size, $\sim 1/n$. Since the expectation values of the transverse operators $\ave{\sigma^{\x}_j}=\ave{\sigma^{\y}_j}=0$ we immediately see that,
\begin{equation}
\widetilde{C}^{(0)}(\sigma_{j_1}^{\x}\sigma_{j_2}^{\x})=\widetilde{C}^{(0)}(\sigma_{j_1}^{\y}\sigma_{j_2}^{\y})=\frac{1}{2} \sin(\pi x_1 )\sin(\pi x_2 ), \quad x_{1,2}=\frac{j_{1,2}-1}{n-1},
\end{equation} 
where we have taken $ x_{1,2}=\frac{j_{1,2}-1}{n-1}$ to conform with previous approaches \cite{Prosenreview, Prosen2, PKS2}. 
One sees then that the transverse 2-point connected correlators do not decay with system size making them truly long-range in the sense of Ref. \cite{longrange}. Since the basis of decoherence is determined by the dissipator Eq. \eqref{maxdriv} to be in the z-direction, this can be understood as a purely quantum effect.

\begin{figure}
 \centering	
\vspace{-1mm}
\includegraphics[width=0.9\columnwidth]{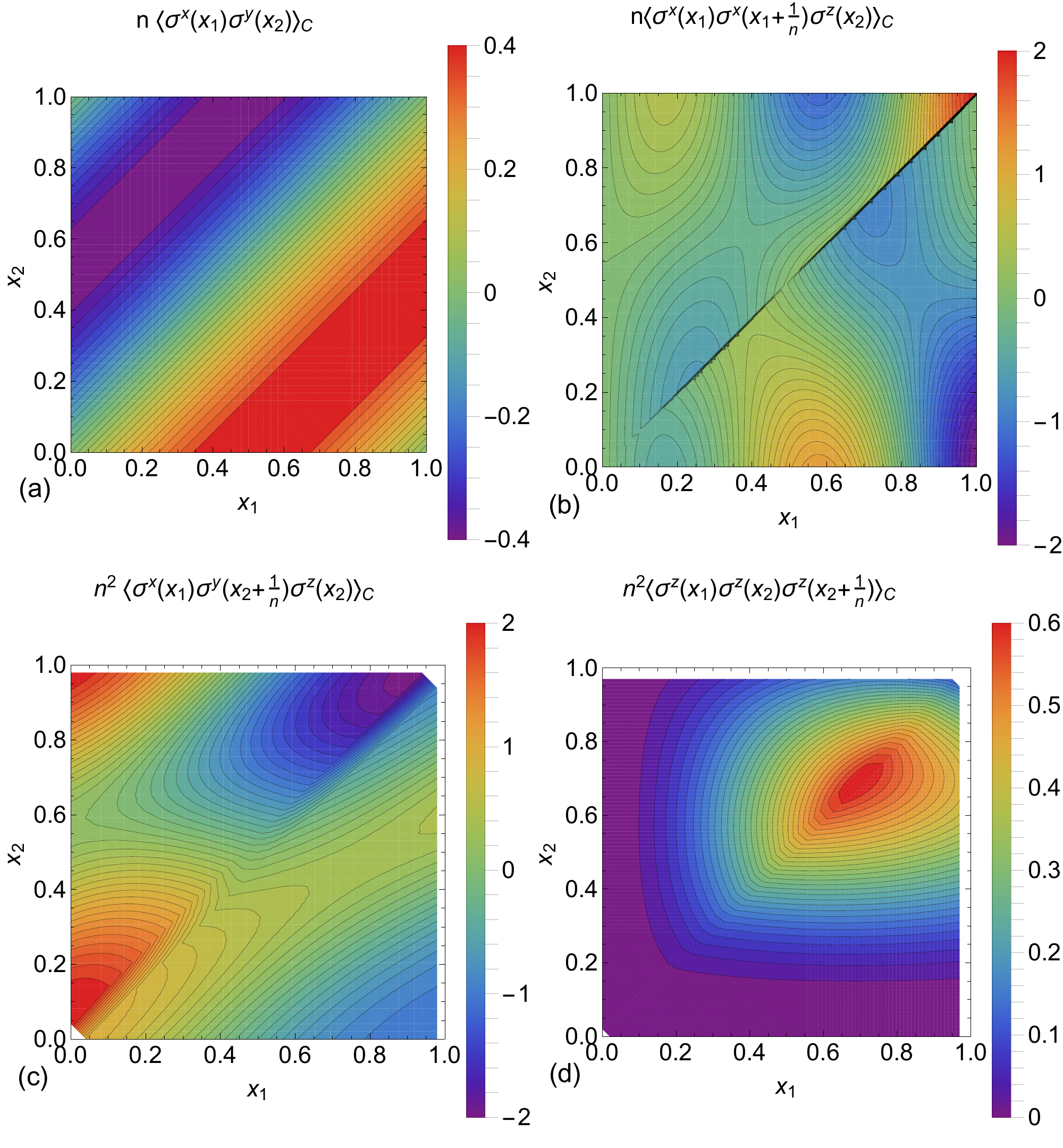}
\vspace{-1mm}
\caption{The two and three-point connected correlation function for $\Delta=1$ and $\epsilon=10$.}
\label{3point}
\end{figure}

Furthermore, when we find the $1/n$ correction we require also (as is given by the explicit representation of the ASOs \cite{Prosen2, Prosenreview}) that the boundary conditions, 
\be 
\frac{\partial \widetilde{C}^{(1)}(\sigma_{j_1}^{\x}\sigma_{j_2}^{\x})}{\partial x_1}\big|_{x_1=1,x_2=1}=0,
\ee 
are satisfied. This fixes uniquely that $\beta=1$, or $\alpha=3/4$ in Eq. \eqref{scale}. Thus,
\begin{align}
\frac{8}{\pi} \widetilde{C}^{(1)}(\sigma_{j_1}^{\x}\sigma_{j_2}^{\x})&=  2 (1 -y_2 ) \cos(\pi y_2) [\pi y_1 \cos(\pi y_1) -\sin(\pi y_1)]-2 y_1 \cos(\pi y_1)\nonumber \\
 &- \pi [y_1(y_1 - 1) + y_2(1 - y_2)] \sin(\pi y_1) \sin(\pi y_2), \label{foxx}
\end{align} 
where, as before, $ x_{1,2}= \frac{j_{1,2}-1}{n-1}$ and $y_1=\min(x_1,x_2)$,  $y_2=\max(x_1,x_2)$.

We find that the mixed transverse 2-point correlator $\widetilde{C}(\sigma_{j_1}^{\x}\sigma_{j_2}^{\y})=C(\sigma_{j_1}^{\x}\sigma_{j_2}^{\y})$ decays with system size as $1/n$ in leading order,
\begin{equation}
\widetilde{C}^{(1)}(\sigma_{j_1}^{\x}\sigma_{j_2}^{\y})=\frac{\pi}{\epsilon}\sin(\pi( x_1-x_2)),\quad \widetilde{C}^{(0)}(\sigma_{j_1}^{\x}\sigma_{j_2}^{\y})=0, \quad x_{1,2}=\frac{j_{1,2}-1}{n-1}.
\end{equation}
The $m$-point connected correlators however become quickly more complicated for higher orders.

The simplest non-zero three point connected correlation function is $\ave{\sigma_{j_1}^{\z}\sigma_{j_2}^{\x}\sigma_{j_3}^{\x}}_c$. It scales as $1/n$ and is given by,
\begin{align}
&\widetilde{C}^{(1)}(\sigma_{j_1}^{\z}\sigma_{j_2}^{\x}\sigma_{j_3}^{\x})=\frac{1}{4} \pi^2 ( x_1-1) \cos(\pi y_3) y_2 \cos(\pi x_1) \cos(\pi y_2)- \nonumber \\ 
&-\frac{1}{4} \pi ( x_1-1) \cos(\pi y_3) \left [ (\cos(\pi x_1) + \pi x_1 \sin(\pi x_1)) \sin(\pi y_2)\right] \nonumber  \\
&- 4 \cos(\pi y_2) \left[ \pi^2 y_2 \cos(\pi x_1)- (2 + \pi^3 x_1 (-1 + y_2)) \sin(\pi x_1)\right]  \nonumber \\
&+ \frac{1}{16\pi}\Big[\left \{ 8 - 5 \pi -2 \pi^3 \left[ (1 - x_1) x_1 +(1 - y_2) y_2+ (1 - y_3) y_3\right] \right \} \cos(\pi x_1) \nonumber  \\
&+ 2 \pi^2 (1 - 2 x_1) \sin(\pi x_1) \sin(\pi y_2) \sin(\pi y_3)\Big], \quad x_1<y_2<y_3,
\end{align}
\begin{align}
&\widetilde{C}^{(1)}(\sigma_{j_1}^{\z}\sigma_{j_2}^{\x}\sigma_{j_3}^{\x})\nonumber \\ 
&=\frac{1}{16 \pi}\Big\{2 \pi \cos(\pi y_3) [-2 \pi^2 x_1 (-1 + y_3) \sin(\pi x_1) \sin(\pi y_2)  \nonumber \\ 
&+\cos(\pi x_1)\cos(\pi y_2) \left \{ 2 \pi^2 y_2 (-1 + y_3) + (-16 + 3 \pi - 2 \pi y_3) \sin(\pi y_2)\right \}]   \nonumber \\ 
&-4 \pi^2 \cos(\pi y_2) (y_2 \cos(\pi x_1)+ \pi x_1(-1 + y_2) \sin(\pi x_1)) \nonumber \\
&- [(8 - 5\pi + 2 \pi^3 \left \{ (-1 + x_1) x_1 + (-1 + y_2) y_2 + (-1 + y_3) y_3)\right \} \cos(\pi x_1) \nonumber \\ 
&+ 2 (1 + \pi^2 (1 -2 x_1))] \sin(\pi x_1) \sin(\pi y_2) \sin(\pi y_3)\Big\}, \quad y_2<x_1<y_3
\end{align}
\begin{align}
&\widetilde{C}^{(1)}(\sigma_{j_1}^{\z}\sigma_{j_2}^{\x}\sigma_{j_3}^{\x}) \nonumber \\ 
&= \frac{1}{4}\pi (1- x_1) \cos(\pi y_3) \left[\pi x_1 \sin(\pi x_1)) \sin(\pi y_2)-\pi y_2 \cos(\pi x_1) \cos(\pi y_2)- (\cos(\pi x_1) \right] \nonumber  \\
&\frac{1}{16\pi} \Big\{- 4 \cos(\pi y_2) \left[ \pi^2 y_2 \cos(\pi x_1)- (2 + \pi^3 x_1 (-1 + y_2)) \sin(\pi x_1)\right]  \nonumber \\
&- \left \{ 8 - 5 \pi -2 \pi^3 \left[ (1 + x_1) x_1 -(1 - y_2) y_2+ (1 - y_3) y_3\right] \right \} \cos(\pi x_1) \nonumber  \\
&- 2 \pi^2 (1 + 2 x_1) \sin(\pi x_1) \sin(\pi y_2) \sin(\pi y_3)\Big\}, \quad y_2<y_3<x_1,
\end{align}
where $ x_{k}= \frac{j_{k}-1}{n-1}$ and $y_2=\min(x_2,x_3)$,  $y_3=\max(x_2,x_3)$.

The simplest four-point function is $\ave{\sigma_{j_1}^{\x}\sigma_{j_2}^{\x}\sigma_{j_3}^{\x}\sigma_{j_4}^{\x}}_c=\ave{\sigma_{j_1}^{\y}\sigma_{j_2}^{\y}\sigma_{j_3}^{\y}\sigma_{j_4}^{\y}}_c$ and is again long range in the sense that it doesn't decay with system size $n$ in the leading order, 
\begin{equation}
\widetilde{C}^{(0)}(\sigma_{j_1}^{\x}\sigma_{j_2}^{\x}\sigma_{j_3}^{\x}\sigma_{j_4}^{\x})=-\frac{3}{8}\sin(\pi x_1)\sin(\pi x_2)\sin(\pi x_3)\sin(\pi x_4), \quad x_{k}=\frac{j_{k}-1}{n-1}.
\end{equation}
\begin{figure}
 \centering	
\vspace{-1mm}
\includegraphics[width=0.95\columnwidth]{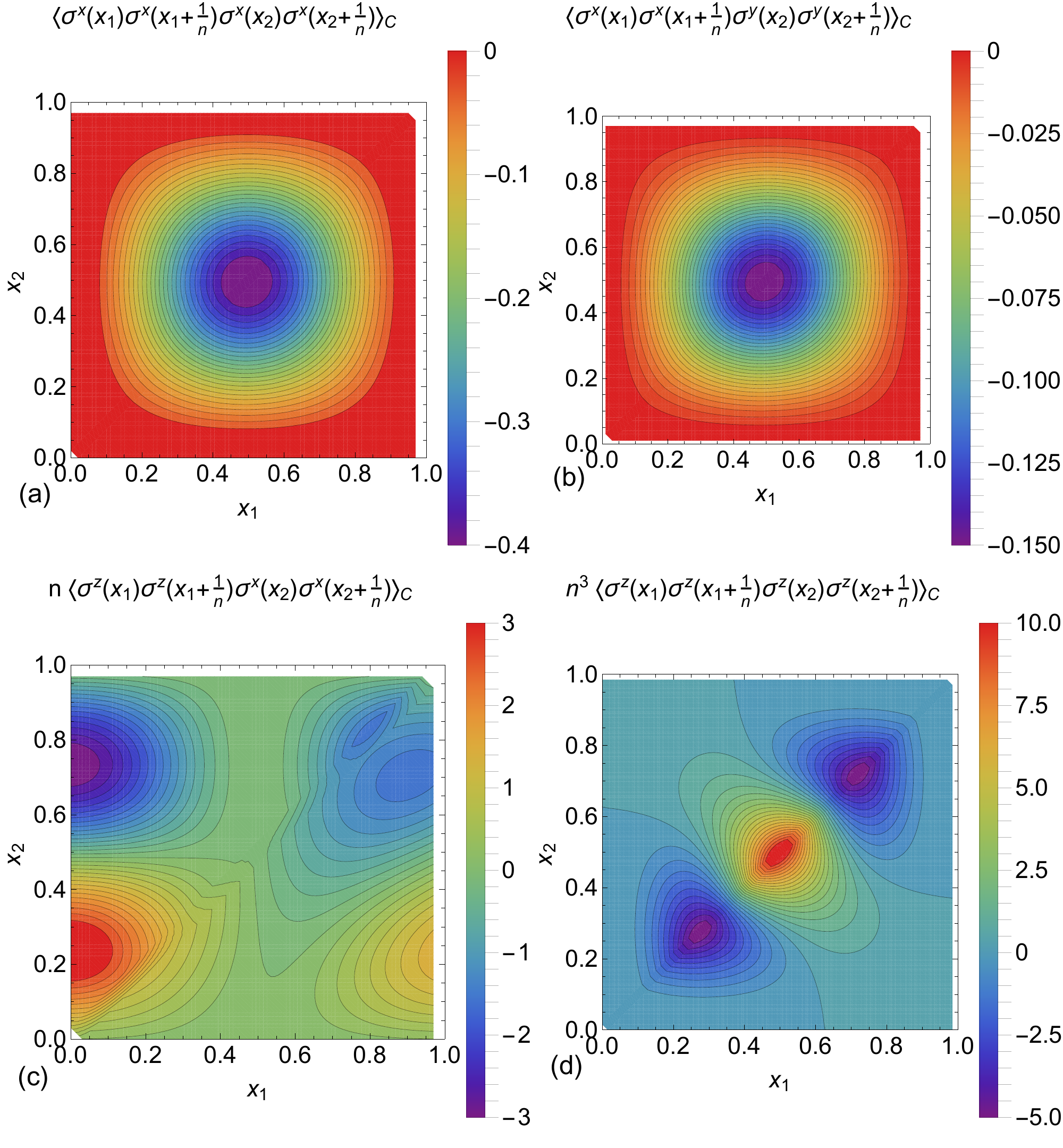}
\vspace{-1mm}
\caption{Some of the 4-point connected correlators calculated at strong coupling $\epsilon=10$ and $\Delta=1$.}
\label{fig1}
\end{figure}
The other correlation functions studied and plotted in Figs. \ref{3point} and \ref{fig1} include $\ave{\sigma_{j_1}^{\z}\sigma_{j_2}^{\z}\sigma_{j_3}^{\z}}_c \propto 1/n^2$, $\ave{\sigma_{j_1}^{\z}\sigma_{j_2}^{\z}\sigma_{j_3}^{\z}\sigma_{j_4}^{\z}}_c \propto 1/n^3$ and $\ave{\sigma_{j_1}^{\x}\sigma_{j_2}^{\x}\sigma_{j_3}^{\z}\sigma_{j_4}^{\z}}_c \propto 1/n$.
Based on our results, although we were unable to prove so in general, we conjecture that in leading order of $1/n$, $m$-point connected correlation functions containing only $\sigma^\z$ operators scale as $1/n^{m-1}$ and that all purely transverse $m$-point connected correlators, i.e., those containing only an even number of $\sigma^\x$ ($\sigma^\y$) do not decay in the leading order, $\sim(1/n)^0$.

\subsection{The easy-plane regime $\Delta<1$}

For $\Delta<1$ only the cubic relation for $\bb{O}_{\rm z}$ is known \cite{IlievskiZunkovic, IlievskiPhD} and it is given by,
\begin{equation}
\kappa_{0}(\gamma,s)(\bb{O}_0\bb{O}_0\bb{O}_{\rm z}+\bb{O}_{\rm z}\bb{O}_0\bb{O}_0)+\bb{O}_0\bb{O}_{\rm z}\bb{O}_0+
\kappa_{1}(\gamma,s)\{\bb{O}_{\rm z},\bb{O}_{\rm z}\}+\kappa_{2}(\gamma,s)\bb{O}_{\rm z}=0,
\label{vtalgdelta}
\end{equation}
where,
\begin{align}
\kappa_{0}(\gamma,s)&=\half-\cos{(2\gamma)},\nonumber \\
\kappa_{1}(\gamma,s)&=1+\cos{(2\gamma)}+\cos{(4\gamma)}-4\cos{(2\gamma s)},\\
\kappa_{2}(\gamma,s)&=12\cos{(2\gamma s)}-2\cos{(4\gamma)}-10-16\cos{(2\gamma)}\sin^{2}{(\gamma s)}+(8-4\cos{(2\gamma s)})[s]^{2}_{q},\nonumber
\end{align}
where $\Delta=\cos(\gamma)$, $q=\exp(\ii \gamma)$, $[x]_q:=(q^x-q^{-x})/(q-q^{-1})$ and $\varepsilon [s]_q=4\ii  \cos(\gamma s)$.

It leads to,  
\begin{equation}
C^{(0)}(\vec{x})|_{\Delta<1}=0, \label{deltasm}
\end{equation}
for correlators containing only $\sigma^\z$. This is incorrect for e.g., 2-point correlators when $x_1=x_2$ since $(\sigma^\z)^2=\one$. This obviously means that there is a discontinuity and hence the assumption that we may exapand in a Taylor series, which we made when deriving Eq. \eqref{set}, is possibly not valid. In fact computing the correlators when the difference between $x_1$ and $x_2$ is of order of lattice spacing using the known discrete solution \cite{Prosen2} shows that the trivial result Eq. \eqref{deltasm} is not valid there. We cannot therefore make any claims near these points, but calculating these correlators for large but finite $n$ shows they are simply constant functions in $x_k$. 
For finite distances among all the $x_k$ and $x_{k'}$ in the continuum limit (meaning that $|j_{k}-j_{k'}| \to\infty$ as $n\to\infty$) the trivial result Eq. \eqref{deltasm} holds. In other words, on the 'large scale' there is no non-smoothness. This is also the reason why the 'large-scale' results also hold for $\Delta=1$ even though there $(\sigma^\alpha_i)(\sigma^\alpha_j)=\one$, when $i=j$, as well. 

Even though we could not find any similar cubic relations for $\bb{O}_+$ nor $\bb{O}_-$ we find that correlators containing these operators behave like those with only $\sigma^\z $, i.e., they are vanishing. In any case, the behavior in this regime is quite trivial as all expectation values of non-local operators decay to 0 exponentially fast, whereas the (ultra)-local expectation values are constant function throughout the chain \cite{Prosenreview}. 

For $\Delta>1$ the non-equilibrium partition function \eqref{scale} decays exponentially fast \cite{Prosenreview, Prosen2} and we can not apply our method as all orders in $1/n$ contribute when calculating the continuum limit Eq. \eqref{set}.

\section{Fluctuations of spin current and magnetization operators} \label{sec:def}

Quantum mechanics is an intrinsically probabilistic theory, therefore in order to achieve a deeper understanding of any physical system we are studying we need to go beyond calculating merely the expectation values of observables we are interested in.

We will study the probability distribution of the total magnetization in the maximally driven open XXZ spin chain, $P(M: =\sum_j \sigma^{\z}_j)$. 

We also study the fluctuations of \emph{the instantaneous spin current} flowing through the system once the system has reached NESS. That is, avoiding certain ambiguities with defining current statistics \cite{ChuchemCohen1}, we want to merely study simply the moments of the instantaneous spin current measured in the long time limit of the system (NESS), i.e., $\ave{j_k ^m}$. 

The local spin current operator is defined via the continuity equation, $\frac{\partial \sigma^{\z}_k}{\partial t}=\ii [H, \sigma^{\z}_k]=j_k-j_{k-1}$, and for the XXZ spin chain is,
\begin{equation}
j_{k}=2 \ii (\sigma^+_k \sigma^-_{k+1}-\sigma^-_k \sigma^+_{k+1}). \label{spincur}
\end{equation}
However, we notice a problem immediately, namely $j_{k} ^m \propto j_{k}$ when $m$ is odd and  $j_{k} ^m \propto ( \one - \sigma^{\z}_k \sigma^{\z}_{k+1})$ when $m$ is even which makes studying this quite trivial.

A possible solution to this is that we may define a multiple site current operator averaged over $K$ sites, similar to the one studied in \cite{KM} and in \cite{RodenWhaley}, $J^{(K)}_k=\sum_{m=k}^{k+K-1} j_{m} / K$. For example, $J^{(1)}_k=j_{k}$, $J^{(2)}_k=(j_{k}+j_{k+1})/2$, etc. The same type of 'space-integrated' current was previously studied in different settings in Refs. \cite{Eisler, Stinchcombe}. 	

These operators physically correspond to the average flow of magnetization between site $k$ and site $K+k$, i.e,
\begin{equation}
\frac{1}{K}\frac{\partial (\sigma^{\z}_k+\sigma^{\z}_{k+1}+ \ldots + \sigma^{\z}_{k+K})}{\partial t}=J^{(K)}_k-J^{(K)}_{k+1}.
\end{equation}
Note that the expectation values $\ave{J^{(K_1)}_k}=\ave{J^{(K_2)}_k}$ for all $K_1, K_2$, which follows from the continuity equation in the long time limit. However, the higher moments are not equal for different $K_1$ and $K_2$. 

Also, these operators still have the property that with their $(K+1)$-th power, $(J^{(K)}_k)^{K+1} \propto J^{(K)}_k$, so we will define an extensive (up to a prefactor of $1/(n-1)$) quantity $J=J^{(n-1)} $, where $n$ is the system size. That is,
\begin{equation}
J=\sum_{k=1}^{n-1} \frac{j_{k}}{n-1}. \label{totalspincur}
\end{equation}
We also define a moment generating function for the total magnetization operator, $M=\sum_{k=1}^{n} \sigma^{\z}_k $ in the steady state, 
\begin{equation}
G_M(\chi)=\ave{{\rm e}^{\ii \chi M}}=\tr ({\rm e}^{\ii \chi M} \rho_\infty). \label{maggen}
\end{equation}

\section{Fluctuations of spin current and magnetization in the maximally boundary driven XXZ spin chain} \label{sec:fluc}

Using the results from Sec.~\ref{sec:xxz} we can study the fluctuations of the spin current and magnetization operators, as defined in the previous section, Sec.~\ref{sec:def}. 
Recall that we defined the averaged spin current operator as $J=\sum_{k=1}^{n-1} \frac{j_k}{n-1}.$ The ASO corresponding to the local spin current $\bb{W}:=\ii( \bb{O_+}\bb{O_-}-\bb{O_-}\bb{O_+)}$ is proportional to $\bb{O}_0$ \cite{Prosen2,IlievskiPhD}, 
\begin{equation}
\bb{W}=-2\ii \left [ s \right]_q \bb{O}_0, \label{propcur}
\end{equation}
where, as before, $\Delta=\cos(\gamma)$, $q=\exp(\ii \gamma)$, $[x]_q:=(q^x-q^{-x})/(q-q^{-1})$ and $\varepsilon [s]_q=4\ii  \cos(\gamma s)$. The fact that $\bb{W} \propto \bb{O}_0$ guarantees the validity of the continuity equation for the magnetization,$\frac{\partial \sigma^{\z}_k}{\partial t}=\ii [H, \sigma^{\z}_k]=j_k-j_{k-1}$, in the long time limit when the system has reached the NESS, i.e., $\frac{\partial \ave{\sigma^{\z}_k}}{\partial t}=0=\ave{j_k}-\ave{j_{k-1}}$ (in other words the expectation values of spin current $j_k$ are equal on all sites $k$). 

We square (take the third power of) Eq. \eqref{totalspincur} and then using the definition Eq. \eqref{spincur}, the properties of the Pauli matrices and the previously mentioned fact that the NESS expectation values $\ave{j_k}=\ave{j_m}$ for all $k,m$ we find (grouping equal terms together), 
\begin{align}
&\ave{J^2}= \label{2nd} \\
&\frac{1}{(n-1)^2}\left( (n-2)(n-3) \ave{j_{1,2}j_{3,4}} +\frac{n-1}{2}-\sum_{k=1}^{n-1}\left( \frac{ \ave{\sigma^{\z}_k\sigma^{\z}_{k+1}}}{2}+2\ave{\sigma^+_k \sigma^-_{k+2}+\sigma^-_k \sigma^+_{k+2}} \right) \right), \nonumber
\end{align}
and,
\begin{align}
\ave{J^3}&=\frac{1}{(n-1)^3}\Big\{- (n-3)(n-4)(n-5)  \ave{j_{1,2}j_{3,4} j_{4,5}}-3(n-2)(n-3) \ave{j_{1,2}}\nonumber \\
&+\ii \sum_k \left (3 \ave{\sigma^+_k \sigma^-_{k+3}-\sigma^-_k \sigma^+_{k+3}}-\ave{\sigma^+_k \sigma^{\z}_{k+1} \sigma^{\z}_{k+2} \sigma^-_{k+3}}+\ave{\sigma^-_k  \sigma^{\z}_{k+1}\sigma^{\z}_{k+2}\sigma^+_{k+3}} \right ) \nonumber \\
&+\sum_k \left ( 3(n-3) \ave{j_{1,2}}\ave{ \sigma^{\z}_k \sigma^{\z}_{k+1}}-12(n-4)\ave{j_{1,2}} \ave{\sigma^+_k \sigma^-_{k+2}+\sigma^-_k \sigma^+_{k+2}}  \right )\Big\}.   \label{3rd}
\end{align}
Let us first consider the isotropic case $\Delta=1$ ($q\to 0$).
The second moment $\ave{J^2}$ contains sums of expectation values of three different types of operators which are non-trivial in the thermodynamic limit $n \rightarrow \infty$: $\ave{j_{1,2}j_{3,4}}$, $\ave{\sigma^{\z}_k\sigma^{\z}_{k+1}}$, and $\ave{\sigma^+_k \sigma^-_{k+2}+\sigma^-_k \sigma^+_{k+2}}$. 
Using Eq. \eqref{propcur} and the asymptotic form of the non-equilibrium partition function Eq. \eqref{scale}, it is easy to show that $\ave{j_{1,2}j_{3,4}} \propto 1/n^4$ in the leading order.

Furthermore we note that the leading order of the other terms cancels exactly. In fact, using, the next to leading order of $\ave{\sigma^{\z}_k\sigma^{\z}_{k+1}}$, which was shown in Ref. \cite{Prosenreview} to be (where, as before $ x_{1,2}=\frac{j_{1,2}-1}{n-1}$),
\begin{align}
& C^{(1)}(\sigma_{j_1}^{\z}\sigma_{j_2}^{\z})= -\frac{\pi}{4} \cos(\pi x_1) \Big[\pi \{(-1 + x_1) x_1 + (-1 + x_2) x_2\} \cos(\pi x_2) \nonumber  \\
 &+ (1 - 2 x_2) \sin(\pi x_2) + \sin(\pi x_1) \{(1 - 2 x_1) \cos(\pi x_2) - 2 \pi x_1 (-1 + y_2) \sin(\pi x_2)\}\Big], \label{sigmazsigmaz}
 \end{align}
together with Eq. \eqref{foxx}, we see that the next-to-leading order $1/n$ cancels as well. However expanding around $x_2=x_1+1/(n-1)$ for large $n$ we are left with a finite contribution of order $1/n^2$.  Similarly for $\ave{J^3}$ we are left with a leading order of $1/n^5$
Finally we find that,
\begin{equation} 
\lim_{n \to \infty } \ave{J^2} = \frac{2 + 5 \pi^2}{8 n^3} \quad \lim_{n \to \infty } \ave{J^3} = \frac{5 \pi^3+8 \pi^2}{\epsilon n^5}
\end{equation}

For a dense set of rational anisotropies, $\Delta=\cos (\pi m/n)$ with $m,n \in \mathbb{Z}$, it is known that the spin current is ballistic (reaching a constant value as $n \to \infty$) as the matrix representation of $\bb{O}_0$ is truncated to a finite dimension. Thus the non-equilibrium partition function is determined by the largest eigenvalue of this truncated transfer matrix in the limit $n \to \infty $ \cite{Prosenreview, Prosen2}. 
Therefore in Eqs. \eqref{2nd} and \eqref{3rd} the terms $ (n-2)(n-3) \ave{j_{1,2}j_{3,4}}$ and $ - (n-3)(n-4)(n-5)  \ave{j_{1,2}j_{3,4} j_{4,5}}$, respectively, will dominate. It is clear that an analogous result holds for all the higher moments. If we let $\lambda_\Delta$ denote the largest eigenvalue of the truncated $\bb{O}_0$ at anisotropy $\Delta$ then from Eq. \eqref{propcur} it follows, 
\begin{equation}
\ave{J^m}=\left ( \frac{-2\ii}{\lambda_\Delta}  \left [ s \right ]_q \right )^m, \label{powerscurdelta}
\end{equation}
where, as always, $\Delta=\cos(\gamma)$, $q=\exp(\ii \gamma)$, $[x]_q:=(q^x-q^{-x})/(q-q^{-1})$ and $\varepsilon [s]_q=4\ii  \cos(\gamma s)$. As $s$ is purely imaginary \cite{IlievskiZunkovic, IlievskiPhD} Eq. \eqref{powerscurdelta} is always real. 

\begin{figure}
 \centering	
\includegraphics[width=1.0\columnwidth]{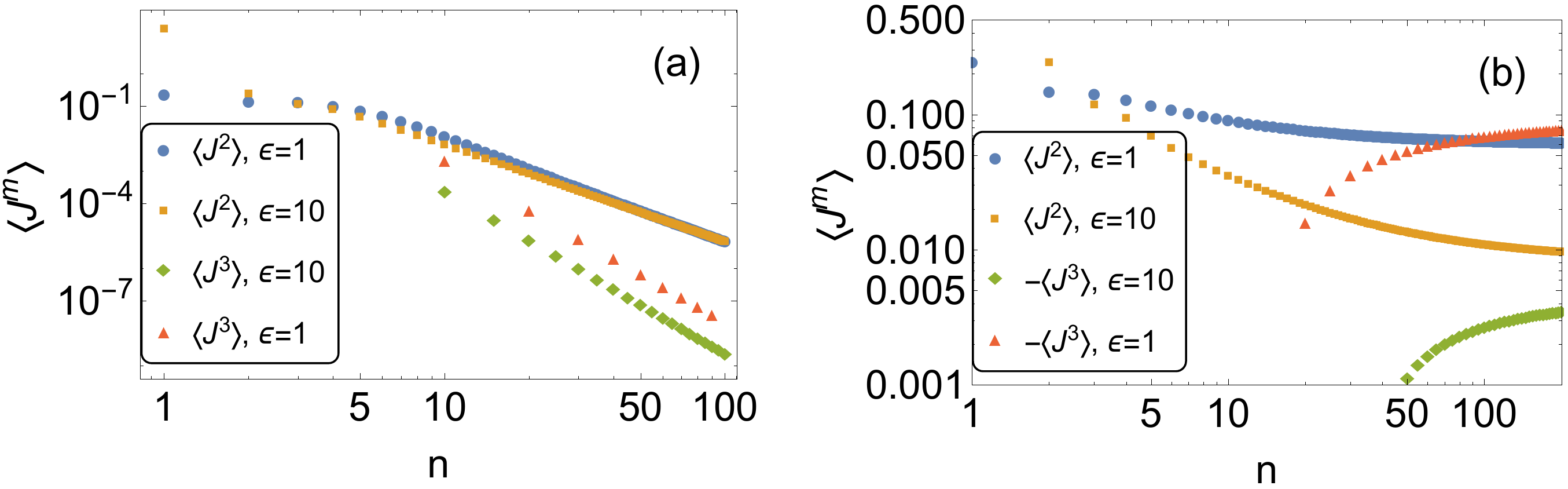}
\vspace{-1mm}
\caption{The current moments for (a) $\Delta=1$ and (b) $\Delta=1/2$ for two coupling parameters $\epsilon=1$ and $\epsilon=10$ obtained by explicit computation of the MPA for the NESS found in Ref. \cite{Prosen2}.}
\label{curfig}
\end{figure}

Finally we note that, from the definition Eq. \eqref{maggen}, Eq. \eqref{exp} and Eq. \eqref{pf}, the moment generating function $G_M(\chi)$ for the total magnetization distribution can be given as 
\begin{equation}
G_M(\chi)= \frac{\bra{0}\bb{G(\chi)}^n\ket{0}}{{\cal{Z}}_{n}}=\frac{\bra{0}(\cos(\chi) \bb{O}_0 + \ii \sin(\chi) \bb{O}_{\rm z}  )^n\ket{0}}{{\cal{Z}}_{n}}. \label{gmmoment0}
\end{equation}

\begin{figure}
 \centering	
\includegraphics[width=1.0\columnwidth]{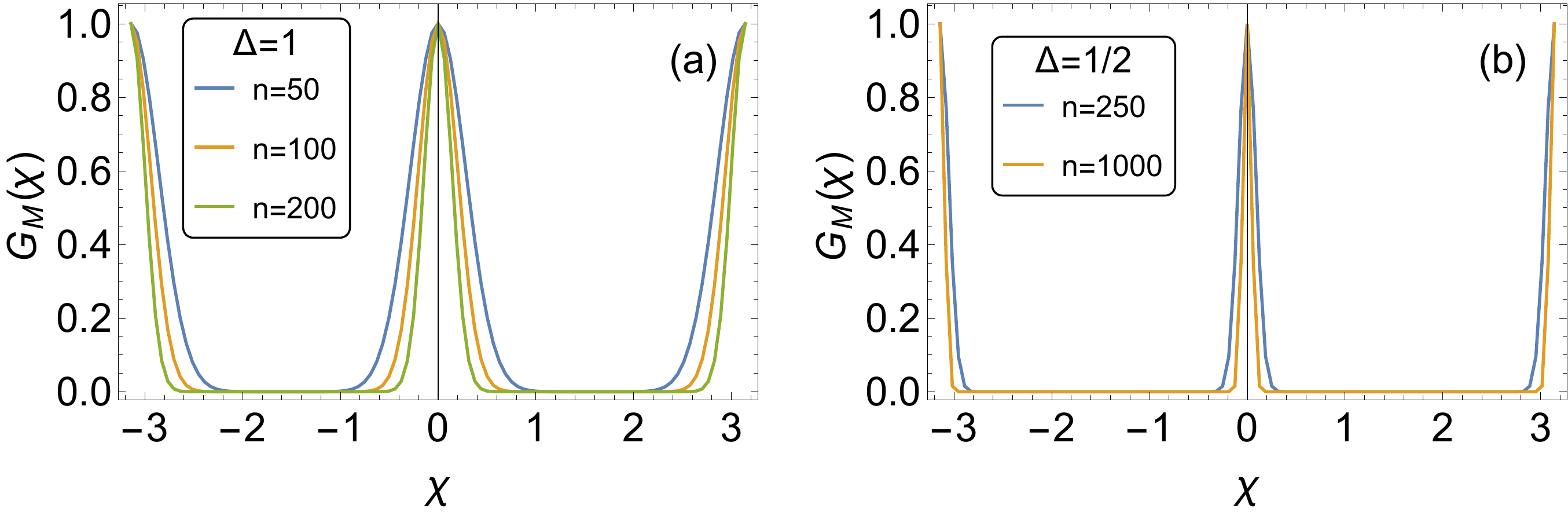}
\vspace{-1mm}
\caption{Asymptotic scaling of the moment generating function of the total magnetization operator for (a) $\Delta=1$, (b) $\Delta=1/2$ obtained by explicit computation of Eq. \eqref{gmmoment0} for the NESS using the MPA found in Ref. \cite{Prosen2} }
\label{magfig}
\end{figure}
We were unable to find closed form expression for the probability distribution of the total magnetization. However, we perform explicit computation of Eq. \eqref{gmmoment0} using the MPA for NESS from Ref. \cite{Prosen2} for large $n$ and plot it in Fig.~(\ref{magfig}), where we notice that the width of the peaks around $k \pi$ seem to scale as $1/\sqrt{n}$. If so, this suggests that if we Fourier transform Eq. \eqref{gmmoment0} before we take the leading order in $1/n$ we find that the probability distribution is actually a Gaussian distribution around $M=0$ with variance $\propto1/\sqrt{n}$. 

\section*{Conclusions}

In this article we have elaborated on a general method to derive all the correlation functions of exactly solvable boundary driven quantum spin chains in the continuum limit. This method is based on solving partial differential equations for scaled correlators one obtains directly provided that a cubic algebra is satisfied by the auxiliary space operators forming a matrix product steady state, and incorporating also the appropriate boundary conditions. We utilise this method to compute various quantities for the exemplar case of the open boundary driven XXZ spin chain \cite{Prosen2}.

More specifically, we computed all the (transverse and longitudinal) spin correlation functions for the boundary driven open isotropic XXX spin chain in the leading order of the continuum limit up to a scaling factor (for which a recursion relation was conjectured based on a known discrete solution \cite{Prosen2}). We derived explicit expressions for certain (up to 4-point) connected correlators and observed that the connected correlators of operators transversal to the basis of decoherence in the $z$-direction (i.e., tensor products of an even number of $\sigma^\x$ and $\sigma^\y$) exhibit long-range order --- that is they do not decay with system size in the leading order. Similar behavior has been previously observed numerically for a related system \cite{longrange}.

Finally, we defined two statistical quantities of interest --- fluctuations of the total spin current (studied previously in Refs. \cite{Eisler, Stinchcombe} in different contexts and also related to a quantity studied in \cite{KM} and in \cite{RodenWhaley}) and the magnetization operators.
We computed the second and third moment of the spin current operator at the isotropic point $\Delta=1$ and all the moments for $\Delta<1$ when $\Delta$ can be expressed via rational multiples of $\pi$ as $\Delta = \cos(\pi n/m)$. At the isotropic point we found that for system size $n$ the second moment decays $\propto 1/n^3$ and the third moment $\propto 1/n^5$. For $\Delta<1$ we find that none of the moments decay with the system size. 

We need to stress that it is possible that the continuum limit is not well-defined when the difference between the operators in the basis for the steady state is of order of the lattice spacing. In this article we focused on the continuum limit of, the previously mentioned, open XXZ spin chain for which a discrete solution is known \cite{Prosen2} and this allowed us to check the validity of the continuum limit.  

Even though our treatment for the continuum limit is exact, it would be interesting to see whether one can find a solution for an open quantum non-equilibrium steady state directly for a continuum system (and thus avoiding knowing an exact solution for any finite, discrete system), either on the level of a quantum field theory or at least perturbatively in lattice spacing in the leading order (in the sense of Eq. \eqref{zero}).

\section*{Acknowledgements}

We thank M. Medenjak and E. Ilievski for useful discussions and acknowledge financial support by Slovenian Research Agency (ARRS), under grants P1-0044, J1-5439 and N1-0025.

\end{document}